\documentstyle[12pt,epsf,rotating,epsfig]{article}
\def\Pom{{I\!\!P}}
\def\lsim{\mathrel{\rlap{\lower4pt\hbox{\hskip1pt$\sim$}}
    \raise1pt\hbox{$<$}}}         %less than or approx. symbol
\def\gsim{\mathrel{\rlap{\lower4pt\hbox{\hskip1pt$\sim$}}
    \raise1pt\hbox{$>$}}}         %greater than or approx. symbol

\newlength{\dinwidth}
\newlength{\dinmargin}
\begin{document}
\noindent
DESY 96--073         \hfill ISSN 0418-9833\\
INP Cracow 1725/PH\\
April 1996      \\ %     \hfill {\bf Draft} ~~~~~~~~~~~~~  \today \\
\begin{center}
  \begin{Large}
{\bf Properties of HERA Events\\ from DIS on Pions in the Proton}\\
  \end{Large}
  \vspace{5mm}
  \begin{large}
M.~Przybycie\'n$^{a}$\footnote{supported by the Polish State
Committee for Scientific Research, grant No.2P03B 244 08p02},
A.~Szczurek$^{b}$, G.~Ingelman$^{cd}$\\
  \end{large}
\end{center}
%  \vspace{3mm}
$^{a}$ Faculty of Physics and Nuclear Techniques, Academy of Mining
and Metallurgy,\\
\hspace*{3mm} Al. Mickiewicza 30, PL-30-059 Cracow, Poland\\
$^{b}$ Institute of Nuclear Physics, ul.Radzikowskiego 152,
PL-31-342 Cracow, Poland\\
$^c$ Deutsches~Elektronen-Synchrotron~DESY,
Notkestrasse~85,~D-22603~Hamburg,~FRG\\
$^d$ Dept. of Radiation Sciences, Uppsala University,
Box 535, S-751 21 Uppsala, Sweden\\
\vspace{5mm}

\begin{quotation}
\noindent
{\bf Abstract.}
Recently the concept of the pion cloud in the nucleon turned out to
be successful in understanding the Gottfried sum rule violation observed
by the New Muon Collaboration and the Drell--Yan asymmetry measured in
NA51 at CERN.  We propose a further possibility to test this concept 
at HERA through the analysis of the structure of deep inelastic scattering (DIS) 
events induced by pion--exchange. Momentum and energy distributions of 
outgoing nucleons as well as rapidity and multiplicity distributions are 
investigated using Monte Carlo simulations. 
Most observables cannot distinguish this process from ordinary DIS, 
but in the energy distribution of final neutrons we find a significantly 
different prediction from the pion cloud model. Forward neutron calorimeters 
will be essential to test the concept of pions in the nucleon.
\end{quotation}

\section{Introduction}

In deep inelastic scattering (DIS) the incident lepton is scattered on 
a coloured quark. Normally this results in a colour field between the struck 
quark and the proton remnant, such that hadrons are produced in the whole 
rapidity region in between. 
In electron-proton collisions at HERA, this leads to particles being produced
also close to the proton beam direction. The recent discovery by 
the ZEUS \cite{lrg_ZEUS} and H1 \cite{lrg_H1} collaborations at HERA
of large rapidity gap events has attracted much interest. 
This new class of DIS events have a large region of forward rapidity 
(i.e. close to the proton beam) where no particles or energy depositions 
are observed.  
The most forward hadronic activity being observed is then actually in the 
central part of the detectors. 
These large rapidity gap events cannot be described by standard models 
for DIS and hadronization \cite{LEPTO,Lund,Werner}. 
Therefore the observation of a surprisingly large fraction ($\sim 10\%$) of 
events with a large rapidity gap strongly suggests the presence of a final 
proton close to the beam momentum. 
These events have, therefore, been primarily interpreted in terms of 
pomeron exchange, although alternative models have recently been proposed 
\cite{SCI}. 

In this interpretation, the lepton interacts with a colorless object having 
the quantum numbers of the vacuum, i.e. the pomeron. The experimental
signature is then a quasi-elastically scattered proton well separated in 
rapidity from the other produced particles. The leading proton escapes 
undetected by the main detector, but may be observed in leading proton 
spectrometers that are coming into operation in both ZEUS and H1. 

In the last few years, experiments on DIS have demonstrated that
the internal structure of the nucleon is more complicated than 
expected. The polarized DIS experiments performed by EMC and SMC at CERN have shown that 
only a small fraction of the proton spin is carried by the valence quarks 
\cite{A89}. In addition, the strong violation of the Gottfried sum rule 
observed by NMC \cite{A91} strongly suggests a $\bar d - \bar u$
asymmetry of the nucleon sea. The new fits of the parton distributions 
\cite{MSR94} to the world deep inelastic and Drell--Yan data (including 
the dedicated NA51 experiment \cite{B94}) seem to confirm the asymmetry.
Both the violation of the Gottfried sum rule and the asymmetry measured
in the Drell--Yan processes can be naturally accounted for by
the presence of pions in the nucleon, as formulated in the pion cloud 
model \cite{pioncloud}.
In view of these successes of the pion cloud model, it is mandatory to consider
its role in other phenomena. 

The presence of such pions leads to an additional mechanism for nucleon 
production in DIS. In fixed target experiments, as
(anti)neutrino deep inelastic scattering \cite{neutrinoDIS} for instance,
it leads to the production of slow protons. The pion-exchange model
describes the proton production on a neutron target \cite{SBD95}
(extracted from deuteron data \cite{BDT94} 
obtained in bubble chamber experiments at CERN). With HERA kinematics
the pion cloud induced mechanism leads to the production of rather
fast forward protons and neutrons. The mechanism is shown schematically in
Fig.~1. The virtual photon `smashes' the virtual colorless pion
into debris and the nucleon (proton or neutron) or an isobar is produced
as a spectator of the reaction. In this respect there is full analogy
to the reaction on the pomeron. Therefore, the pion cloud
induced mechanism could also lead to rapidity gap events. 
In this paper, we investigate this processes and present quantitative 
results, not previously available in the literature. 

To understand these processes, not only protons but also neutrons
in the forward directions are interesting \cite{LF92,HLNSS94}.
Recently the ZEUS collaboration has installed a forward neutron
calorimeter (FNCAL) \cite{FNC} which will provide additional experimental
information. In analogy to the hadronic reaction
$pp\to nX$, the pion-exchange is expected to be the dominant mechanism of
the fast neutron production also at HERA 
\cite{LF92,HLNSS94}. Thus, HERA open new possibilities to test the concept 
of pion exchange and the pionic structure in the nucleon.

In the present paper we study several quantities which could be analyzed
in HERA experiments; in particular using the main calorimeter, the 
leading proton spectrometer \cite{LPS} (LPS) and the forward neutron 
calorimeter \cite{FNC} in ZEUS. The main aim of the study is to find the best
signal to identify the discussed mechanism of scattering on a pion in 
the proton.

\begin{figure}[t]
\epsfig{file=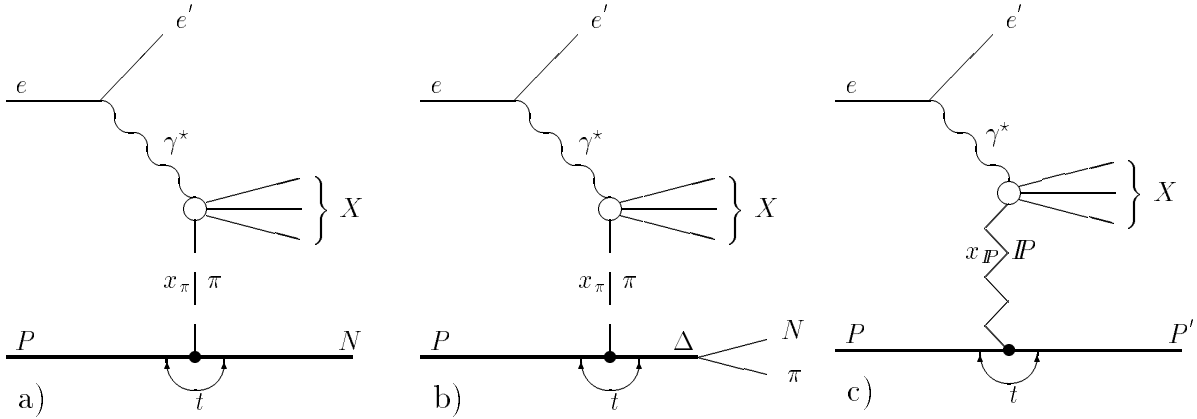,
width=21cm,bbllx=15pt,bblly=280pt,bburx=700pt,bbury=480pt,clip=}
\caption{\it Fast forward nucleon production at HERA: 
(a) direct production through pion-exchange, 
(b) indirect production via a $\Delta$ resonance in pion-exchange,
(c) pomeron-exchange.}
\end{figure}

\section{Pion--exchange mechanism of fast nucleon production}

In the meson cloud model \cite{HSS94} the nucleon is viewed as a quark
core (called a bare nucleon) accompanied by the mesonic cloud.
Restricting to the most important $\pi N$ component, the Fock
state decomposition of the light-cone proton is
\begin{eqnarray}
|p\rangle={\sqrt{Z}}\Big[|(3q)\rangle
   + \int dy\,d^2\vec k_T\phi(z,\vec p_T)\Big(
     \sqrt{1\over 3} |p\pi^0,z,\vec p_T\rangle
   + \sqrt{2\over 3} |n\pi^+,z,\vec p_T\rangle\Big)+\,... \,\Big],
\label{WFp}
\end{eqnarray}
with $Z$ being the wave function renormalization constant which can be
calculated by imposing the normalization condition $\langle p|p\rangle=1$.
$\phi(z,\vec p_T)$ is the light cone wave function of the $\pi N$ Fock state,
where $z$ is the longitudinal momentum fraction of the bare nucleon and
$\vec p_T$ its transverse momentum.

The presence of virtual pions in the nucleon leads to an additional
mechanism for nucleon production referred to as `direct spectator' (Fig.~1a) 
and `sequential spectator' (Fig.~1b) processes. The pion
in the nucleon interacts with a virtual $\gamma$ producing 
a system $X$. For comparison we show the pomeron-exchange
mechanism in Fig.~1c. The cross section for the semi-inclusive spectator
process $ep\to e'NX$ can be written as
\begin{equation}
\frac{d^4 \sigma^{\rm sp} \bigl( ep\to e'NX \bigr) }
{dx dQ^{2} dz dp_{T}^2}
=  \frac{1}{z} f_{\pi N}(1-z,t) \frac{d \sigma^{e \pi}(x/(1-z)) }
{d(x/(1-z))\,dQ^{2}}
\; .
\end{equation}
The presence of the $\pi \Delta$ Fock component in the proton leads to
the production of a spectator $\Delta$ which decays into a pion and
a nucleon. The one-pion exchange contribution to the inclusive cross
section can be obtained by integrating over unmeasured quantities
\begin{equation}
\frac{ d\sigma^{e p}(x,Q^{2}) } {dx \, dQ^2} =
\int_{0}^{1-x} {dz} \int_{-\infty}^{t(0,z)} dt
\;  f_{\pi N}(1-z,t)
\frac{d\sigma^{e \pi}(x/(1-z),Q^2)} {d(x/(1-z)) \, dQ^2} ,
\end{equation}
where  $\sigma^{e \pi}$ is the cross section for the inclusive deep
inelastic scattering of the electron from the virtual pion. In practical
calculations the on-mass-shell $e \pi$ cross section can be used.

The probability density to find a meson with light-cone momentum fraction
$x_{\pi} = (1-z)$ and four-momentum squared $t$ 
(or alternatively transverse momentum $ p_{T}^2= -t(1-x_\pi)-m^2x_{\pi}^2$) 
is referred to as the splitting function, which quantifies the presence
of virtual mesons in the nucleon.
The splitting function $f(x_\pi,t)$ to the $\pi N$ Fock state (Fig.~1a) is
\begin{equation}
f_{\pi N} (x_\pi,t) = \frac{3 g_{p \pi^0 p}^2}{16\pi^2}
x_\pi \frac{ (-t) |F_{\pi N}(x_\pi,t)|^2 } {(t - m_{\pi}^2)^2} ,
\label{splitpiN}
\end{equation}
and to the $\pi \Delta$ Fock state (Fig.~1b) is 
\begin{equation}
f_{\pi \Delta} (x_\pi,t) = \frac{2 g_{p \pi^- \Delta^{++}}^2}{16\pi^2}
x_\pi \frac{ (M_{+}^2 - t)^2 (M_{-}^2 - t) |F_{\pi \Delta}(x_\pi,t)|^2 }
{ 6 m_N^2 m_{\Delta}^2 (t - m_{\pi}^2)^2 } ,
\label{splitpiDelta}
\end{equation}
where $M_{+} = m_{\Delta} + m_N$ and $M_{-} = m_{\Delta} - m_N$ .
The couplings $g^2$ depend on the process, but via the isospin relations 
$g^2_{p\to \pi^+ n} : g^2_{p\to \pi^0 p}=2:1$ and $g^2_{p\to \pi^+ \Delta^0} : 
g^2_{p\to \pi^0 \Delta^+} : g^2_{p\to \pi^- \Delta^{++}}=1:2:3$ 
there are only two independent couplings which we take as 
$g^2_{p\to \pi^0 p}/4\pi = 13.6$ \cite{trs91} 
and $g^2_{p\to \pi^- \Delta^{++}}/4\pi = 12.3$ GeV$^{-2}$ \cite{hhs89}.
The $F_{MB}(x_\pi,t)$ are vertex form factors, which account for the extended
nature of the hadrons involved. The form factors used in meson exchange
models are usually taken to be functions of $t$ only.
As discussed in ref.~\cite{HSS94}
such form factors are a source of momentum sum rule violation and 
it was therefore suggested to use form factors which are 
functions of the invariant mass of the intermediate meson-baryon system,
i.e. $M_{MB}^2(x_\pi,p_{T}^2)= \frac{m^2_\pi +p_{T}^2}{x_\pi}
+\frac{m_B^2 +p_{T}^2}{1-x_\pi}. $

It can be shown that such a vertex function arises naturally
if one computes the splitting function $f(x_\pi,t)$ in time-ordered
perturbation theory in the infinite momentum frame \cite{HSS93}.
This functional form is typical for parameterizing
the light-cone wave function of composed systems (see e.g. \cite{lfcqm}).

In all calculations discussed below
the vertex form factors have been assumed in the exponential form
\begin{equation}
F_{MB}(x_\pi,p_{T}^2) =
\exp \left[- \frac{M_{MB}^2(x_\pi,p_{T}^2) - m_N^2}
{2\Lambda_{MB}^2} \right].
\label{formfac}
\end{equation}
By using the kinematical relation \cite{S72}
\begin{equation}
t = {-p_{T}^2 \over 1-x_\pi} - x_\pi
({m_{B}^2 \over 1-x_\pi} - m_{N}^{2})
\end{equation}
the form factor given by Eq.(\ref{formfac}) can be equivalently
expressed in terms of $x_\pi$ and $t$ in the simple form:
\begin{equation}
F_{MB}(x_\pi,t) =
 \exp \left[ - {m_{\pi}^{2} - t \over 2 \Lambda^{2}_{MB} x_\pi} \right] \; .
\label{formfact}
\end{equation}
The cut-off parameters used in the present calculation
($\Lambda_{\pi N}$ = 1.10 GeV and $\Lambda_{\pi \Delta}$ = 0.98 GeV)
have been determined from the analysis of the particle spectra
for high-energy neutron and $\Delta$ production \cite{HSS94}, i.e.
$pp\to nX$ and $pp\to \Delta^{++}X$. 
With these cut-off parameters the NMC result for the Gottfried sum rule
\cite{A91} which depends sensitively on $\Lambda_{MB}$, has been
reproduced \cite{HSS94}. Furthermore the model describes the
$\overline{u}-\overline{d}$ asymmetry extracted recently from
the Drell-Yan experiment NA51 at CERN \cite{DYMCM}.
We note, however, that all results of this paper would be quite similar 
if traditional dipole form factors with cut-off parameter of 
1.0--1.2 GeV had been used instead of Eq.~(\ref{formfac}).

In hadronic reactions quite often the Regge approach was used rather 
than the light--cone approach. In order to obtain the flux factor in 
the Regge approach it is sufficient to replace in Eq.(\ref{splitpiN})
$x_{\pi}$ by $x_{\pi}^{1-2\alpha_{\pi}(t)}$, where the pion's Regge 
trajectory $\alpha_{\pi}(t)=\alpha_{\pi}^{'}(t-m_{\pi}^{2})$. The 
reggeization is important for small $x_{\pi}$ and/or large $t$.
This is a kinematical region where the flux factor, especially with the
vertex form factor Eq.~(\ref{formfac}), is rather small.
Furthermore in the Regge approach, in contrast to the light--cone 
approach, it is not clear whether it would be fully consistent in the
lepton DIS to use the on--shell pion structure function.
However, since the difference is important only in very limited  region 
of the phase space, in practice both approches lead to almost identical 
flux factors.

%---------------------------------------------------------------------

\section{Results and Discussion}

The formalism presented above has been implemented in the Monte
Carlo program {\sc Pompyt} version 2.3 \cite{Pompyt}. This program, which 
was originally for diffractive interactions via pomeron exchange, 
simulates the interaction dynamics resulting in the complete final 
state of particles. 
The basic hard scattering and perturbative QCD parton emission processes
are treated based on the program {\sc Pythia} \cite{Pythia} and the 
subsequent hadronization is according to the Lund string model \cite{Lund} 
in its
Monte Carlo implementation {\sc Jetset} \cite{Pythia} which also handles 
particle decays. 

The main difference in comparison to the pomeron case is the replacement
of the pomeron flux factor by the pion flux factors given by
Eqs.~(\ref{splitpiN},\ref{splitpiDelta}) and the pomeron structure
function by the pion structure function. 
The pion case is better contrained than the pomeron case, due to the 
better known pion structure function where those for the on-shell pion 
can be used. The pion parton densities from the parametrisation 
GRV-P HO ($\overline{MS}$) \cite{GRV92_pi} is therefore used.
It is important to mention in
this context that the absolute normalization of the cross section for
the production of the spectator nucleon via pion-exchange mechanism
depends on the absolute value of the pion structure function. 
At the small-$x$ relevant at HERA, the structure function is
completely dominated by the pion sea contribution which is not very well
known. Experimentally the pion structure function can be determined from
the Drell--Yan processes only for $x > 0.1$ \cite{B83,SMRS92}.
If the pion-exchange mechanism is the dominant mechanism of fast neutron
production, the coincidence measurement of scattered electrons and
forward neutrons may allow
the determination of the pion deep inelastic structure function
\cite{HLNSS94}. When considering the event structure, however, the
precise value of $F_{2}^{\pi}$ is not required. 

When the deep inelastic scattering is on a valence quark (antiquark)
the pion remnant is simply the remaining antiquark (quark). 
A colour triplet string is then stretched between them and hadronization 
described with the Lund string model \cite{Lund}.
In case it is a sea quark (antiquark) that was struck, 
the pion remnant contains the associated sea antiquark (quark) in
addition to the valence quark and antiquark. A string is then stretched
between the struck quark (antiquark) and a valence antiquark (quark), whereas
the remaining valence quark (antiquark) forms a meson together with 
the spectator sea antiquark (quark).

For the results presented below we have made simulations corresponding to 
the HERA conditions, i.e. $26.7\: GeV$ electrons on $820\: GeV$ protons. 
The results for the above pion exchange mechanism are compared with normal
DIS on the proton, which is simulated with {\sc Lepto} 6.3 \cite{LEPTO}
using the MRS(D-') parton distributions \cite{MRS93}. 
In all cases, events are simulated according to the cross section formulae
and are constrained to be in the kinematical region 
$x>10^{-5}$, $Q^2>4\: GeV^2$. 

In Fig.~2 we show the resulting energy spectra of nucleons 
($p,\bar{p},n,\bar{n}$) in the lab frame of HERA. This is of direct interest
for measurements in the leading proton spectrometer \cite{LPS} and 
forward neutron calorimeter \cite{FNC}.
Neutrons from the pion exchange mechanism have large energies giving a 
spectrum with a broad peak around $E\approx 0.7E_{beam}$, 
i.e. around 500 $GeV$, 
whereas the corresponding spectrum from DIS on the proton decreases 
monotonically with increasing neutron energy. In the region of interest,
say 400--700 $GeV$, the two processes have a similar absolute magnitude. 
An observable effect from DIS on a pion should be therefore possible. 

\begin{figure}[t]
\epsfig{file=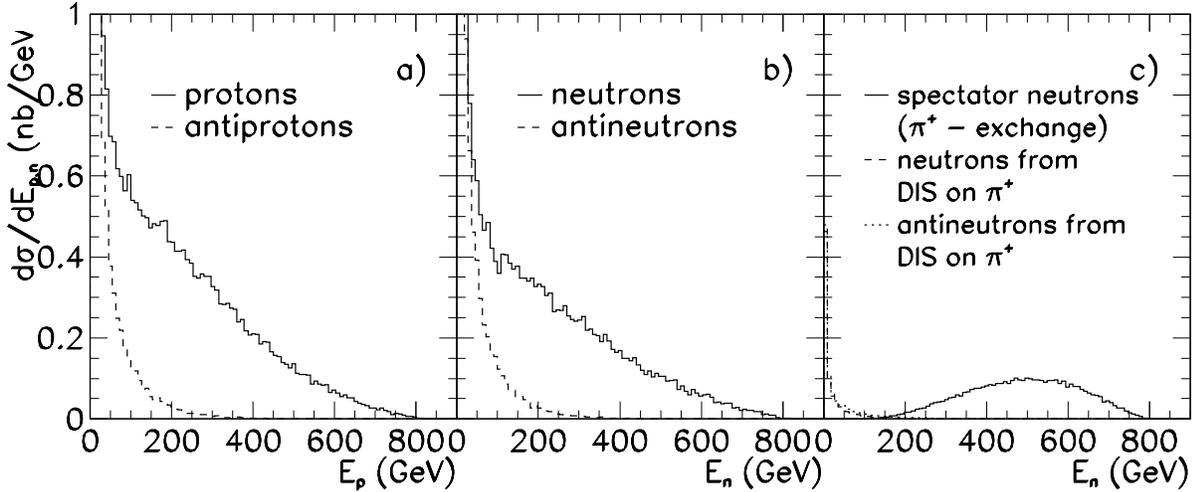,width=16.5cm,bbllx=30pt,
bblly=300pt,bburx=550pt,bbury=500pt,clip=}
\caption{{\it 
Energy spectra in the HERA lab frame for nucleons ($p,\bar{p},n,\bar{n}$) from
(a,b) conventional DIS on a proton (obtained with {\sc Lepto}) 
and (c) DIS on an exchanged $\pi^+$ (obtained with {\sc Pompyt}).}}
\end{figure}

While the energy distribution of primary $\Delta$'s is very similar
to that of the direct neutron production \cite{HSS94},
after the $\Delta \rightarrow n \pi$ decay the energy distribution of
the secondary nucleons becomes peaked at smaller energies of about
400 GeV \cite{HNSSZ95}.
The two-step mechanism is, however, much less important for the
production of neutrons. First of all the probability of the $\pi \Delta$
Fock states in the light-cone nucleon wave function is much smaller than
the probability of the $\pi N$ component:
$P_{\pi \Delta} \approx 0.3 P_{\pi N}$ \cite{HSS94}.
Secondly, the isospin Clebsch-Gordan coefficients favour
the decay of the $\Delta$ into the proton over the decay into
the neutron channel with the proton/neutron branching ratio 
${7 \over 9} : {2 \over 9}$. The analogous branching ratio for the
direct component is ${1 \over 3} : {2 \over 3}$.
All this imply that both the 1-step and 2-step
mechanisms produce comparable amounts of protons. In contrast,
the two--step mechanism produces about 10 times less neutrons than the
1-step mechanism. This means that in a first approximation the
two-step process may be neglected for the spectrum of neutrons.
Therefore we concentrate on the comparison of DIS on $\pi^{+}$, 
having a neutron as spectator, with standard DIS on the proton.

The calculated transverse momentum ($p_T$) distributions are shown in Fig.~3.
As can be seen, the distribution of the spectator neutrons
falls faster with increasing $p_T^2$ than that from standard DIS.
It can be expected that the distribution of neutrons produced in
the two-step process in Fig.~1b is less steep than those produced 
in the direct process in Fig.1a.
The higher overall level of DIS on the proton can be reduced by a cut in 
neutron energy, as is obvious from Fig.~2. Still, the difference in shape 
of the $p_T^2$-spectra in Fig.~3 is presumably too small to be exploited 
experimentally. A safe conclusion does, however, require further analysis
including, e.g., finite angular acceptance of FNCAL.

\begin{figure}[t]
\epsfig{file=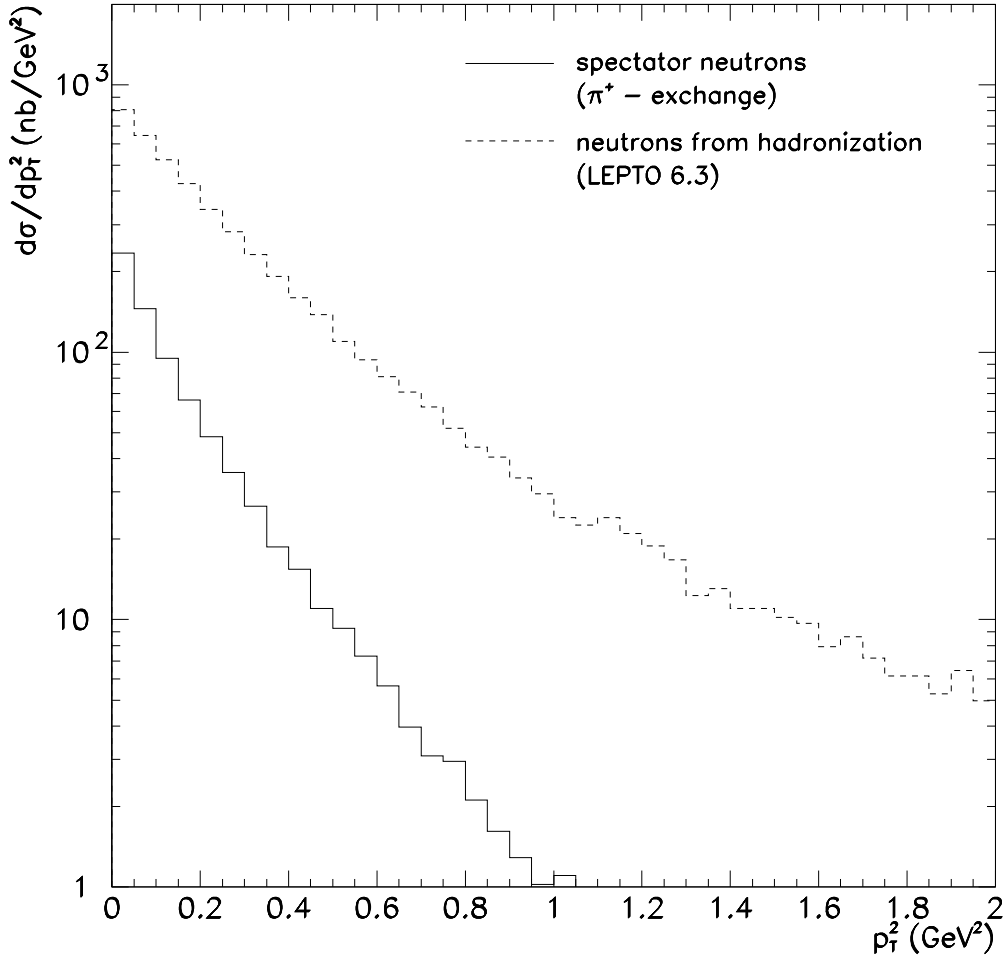,width=15cm,bbllx=-20pt,
bblly=150pt,bburx=550pt,bbury=650pt,clip=}
\caption{{\it 
Distribution in transverse momentum of neutrons from DIS on
the proton (dashed histogram from {\sc Lepto}) and from DIS on
a $\pi^{+}$ with a neutron as a spectator 
(solid histogram from {\sc Pompyt}).}}
\end{figure}
\begin{figure}[t]
\epsfig{file=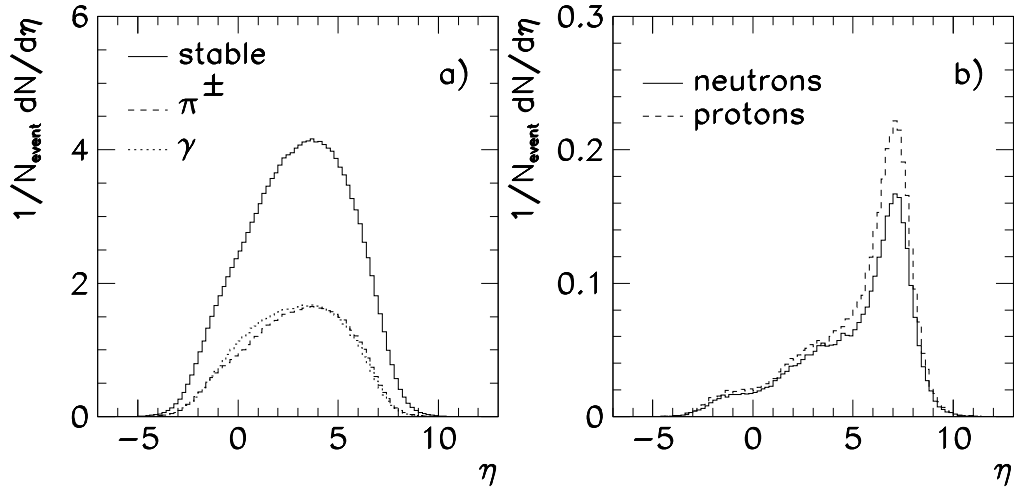,width=15cm,
bbllx=0pt,bblly=280pt,bburx=550pt,bbury=520pt,clip=}
\caption{\it 
Rapidity distributions of (a) all stable particles, charged
pions and $\gamma$'s and (b) protons and neutrons from DIS on the proton 
as obtained from {\sc Lepto}.}
\end{figure}

To study other characteristics of events arising through DIS 
on a virtual pion and compare with standard DIS on the proton, 
we consider spectra of different quantities normalized as
\begin{equation}
f(\kappa) \equiv {1 \over N_{event}} {dN \over d\kappa} \; ,
\end{equation}
where $\kappa$ can be any kinematical variable
and $N_{event}$ is the number of events. 
This give emphasis to shapes irrespectively of normalisation and statistics
(of data and Monte Carlo samples). 

A quantity with especially nice transformation properties under 
longitudinal boosts is rapidity defined as
\begin{equation}
y = \frac{1}{2} ln \left( {E + p_{z} \over E - p_{z}} \right) \; ,
\label{rapidity}
\end{equation}
where $E$ is the energy and $p_{z}$ the longitudinal momentum along the proton 
beam axis. For massless particles this quantity is identical to the 
pseudo-rapidity defined by 
\begin{equation}
\eta = -ln \, tan({\theta/2}) \;
\label{pseudo}
\end{equation}
where $\theta$ is the angle of a particle with respect to the proton beam,
i.e. $\eta >0$ is the proton hemisphere in the HERA lab frame.

In Fig.~4 and 5 we show the pseudo-rapidity distributions of different 
particle species produced in DIS on the proton and DIS on a $\pi^+$, 
respectively. 
In Fig.~5a spectator neutrons are not included, but shown separately in 
Fig.~5b.
For example, the size of the beam pipe hole in FCAL ($\theta=1.5^{0}$),
assures that in almost 100\% of the spectator nucleons (proton/neutron)
leaves the main ZEUS detector without any energy loss.
As seen by comparing Fig.~4a and Fig.~5a the pseudo-rapidity spectra
of $\pi^\pm$ and $\gamma$ are rather similar in the two cases.
The pseudo-rapidity spectrum of spectator neutrons (Fig.5b) has a
maximum at only a slightly higher value compared to the peak of neutrons 
from non-diffractive DIS on the proton (Fig.~4b). 
These predicted neutron distributions should be considered in the 
context of the pseudo-rapidity coverage of the forward neutron calorimeter.
In general, the neutron acceptance is a complicated function of both polar
and azimuthal angle. The ZEUS FNCAL geometry limits pseudo-rapidity coverage
approximately to $7\lsim \eta \lsim 10$.
The Lund hadronization model
predicts a small amount of nucleon-antinucleon pairs produced
in DIS on the pion (Fig.5cd). 

\begin{figure}[tb]
\epsfig{file=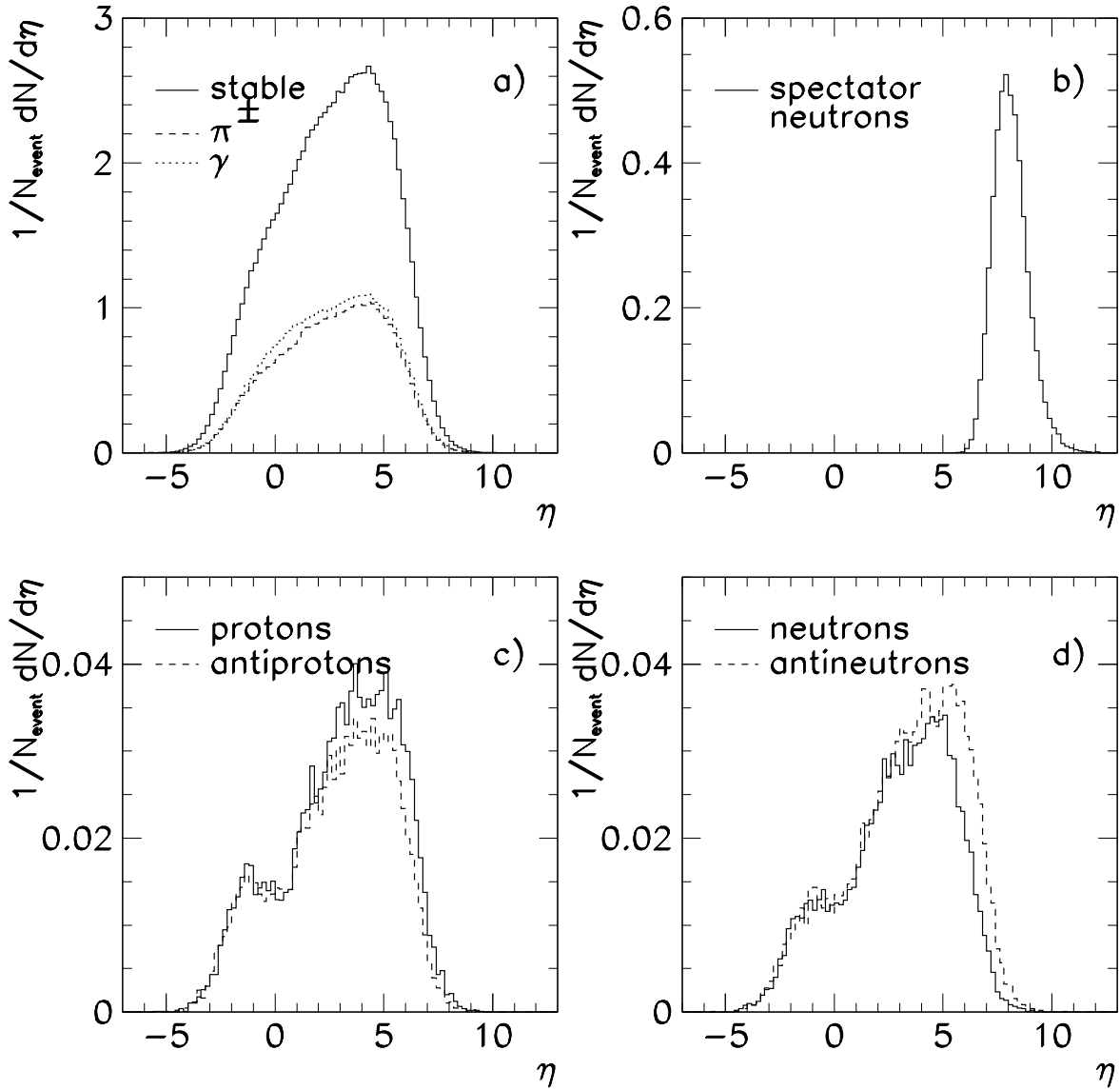,width=16.5cm,bbllx=20pt,
bblly=150pt,bburx=550pt,bbury=650pt,clip=}
\caption{\it 
Rapidity distributions of the specified particles produced in DIS on a 
$\pi^+$ (neutron spectator) as obtained from {\sc Pompyt}.}
\end{figure}

The pseudo-rapidity variable is of particular interest in the context of large
rapidity gap events. These have been defined by $\eta_{max}$ giving, 
in each event, the maximum pseudo-rapidity where an energy deposition is 
observed. Based on our Monte Carlo simulated events using {\sc Lepto} and 
{\sc Pompyt} we extract this 
$\eta_{max}$-variable and show its distribution in Fig.~6 for conventional 
non-diffractive DIS on the proton, DIS on an exchanged $\pi^+$ and diffractive
DIS on a pomeron. 
Since our aim here is to demonstrate the genuine physics effects 
of the models, we have not included any experimental acceptance effects 
or rapidity gap requirements in this study. Doing this will severely distort 
the distributions at large $\eta_{max}$ and, therefore, one cannot make 
direct comparisons with the available measured distributions. 
Thus, from this model study we find a shift of about one unit towards 
smaller $\eta_{max}$ in case of DIS on the pion as compared to normal DIS.
For $\eta_{max}\lsim 6$, these two processes contribute about equally to 
the rate.
\begin{figure}[tb]
\epsfig{file=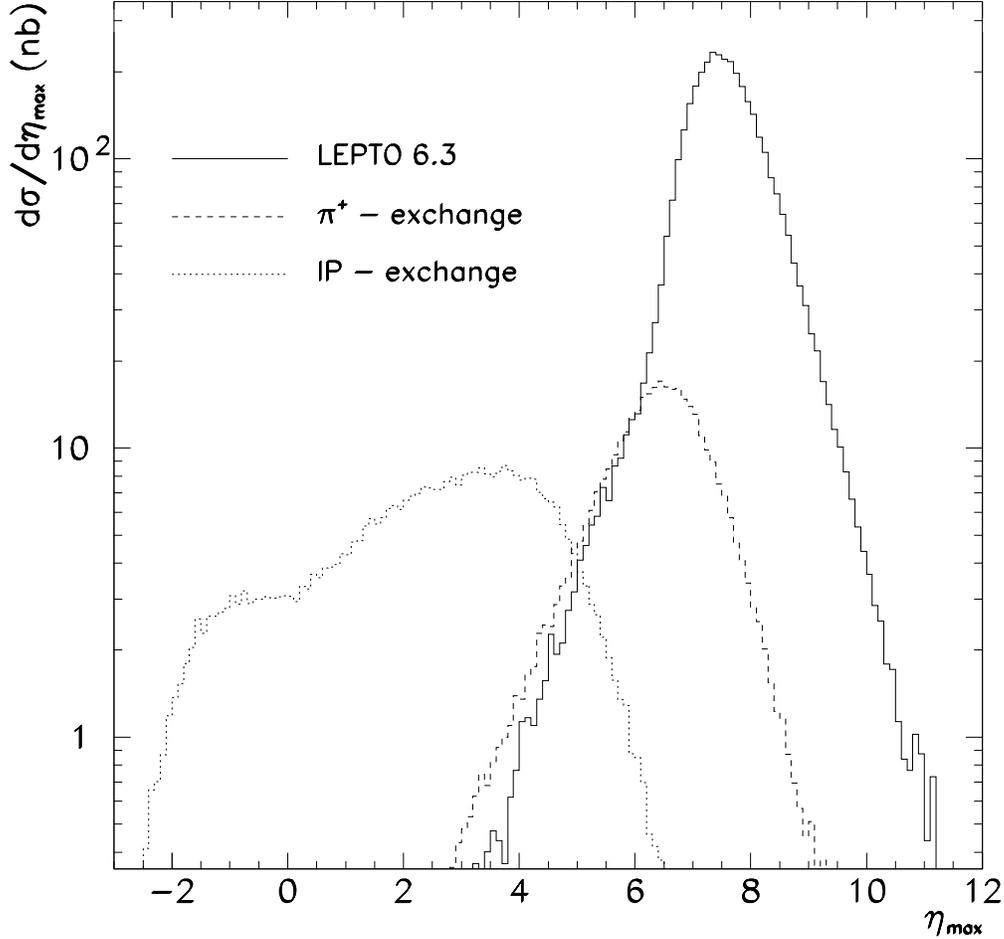,width=15cm,bbllx=-30pt,
bblly=150pt,bburx=550pt,bbury=630pt,clip=}
\caption{\it 
Distribution of $\eta_{max}$ (see text) in non-diffractive DIS on
the proton (solid), in DIS on the virtual $\pi^{+}$ (dashed)
and in DIS on the pomeron (dotted); pure physics of the models without 
experimental acceptance effects.}
\end{figure}

For the spectrum of $\eta_{max}$ for DIS on the pomeron, we have taken
a set of parameters which is usually called `hard pomeron' in the literature. 
The pomeron is assumed to contain equal amounts of the light quarks, i.e.
$u = \bar{u} = d = \bar{d}$, each with a density distribution 
\begin{equation}
z q(z) = {6 \over 4} z (1-z) \; ,
\end{equation}
with the normalization chosen such that the parton distributions
fulfill the momentum sum rule.
The pomeron flux factor is here taken as the ratio of the
single diffractive cross section and the pomeron-proton total cross
section \cite{IS85}
\begin{equation}
f_{\Pom/p}(x_{\Pom},t) =
 \frac{d \sigma / dx_{\Pom} dt}{\sigma(\Pom p \rightarrow X)} =
{1 \over 2.3} {1 \over x_{\Pom}} (3.19 e^{8t} + 0.212 e^{3t}) \; ,
\end{equation}
where the simple parameterization is obtained by fitting the numerator
to single diffractive cross section and the denominator is taken as
$\sigma (\Pom p \rightarrow X) = 2.3 \; mb$ obtained from a Regge
analysis \cite{BCSS87} of elastic and single diffractive scattering.
The resulting $\eta_{max}$ distribution from DIS on the pomeron is considerably 
different from the other two cases. 

From Fig.~6 one may conclude that the pion--exchange induced DIS 
leads to events with intermediate size rapidity gaps rather than to
those with large gaps. 
Nonetheless it is important to verify experimentally the effect of the 
pion exchange by, e.g., correlating $\eta_{max}$ with fast forward
neutrons measured in FNCAL (for technical details see \cite{Brk95}).
\begin{figure}[tb]
\epsfig{file=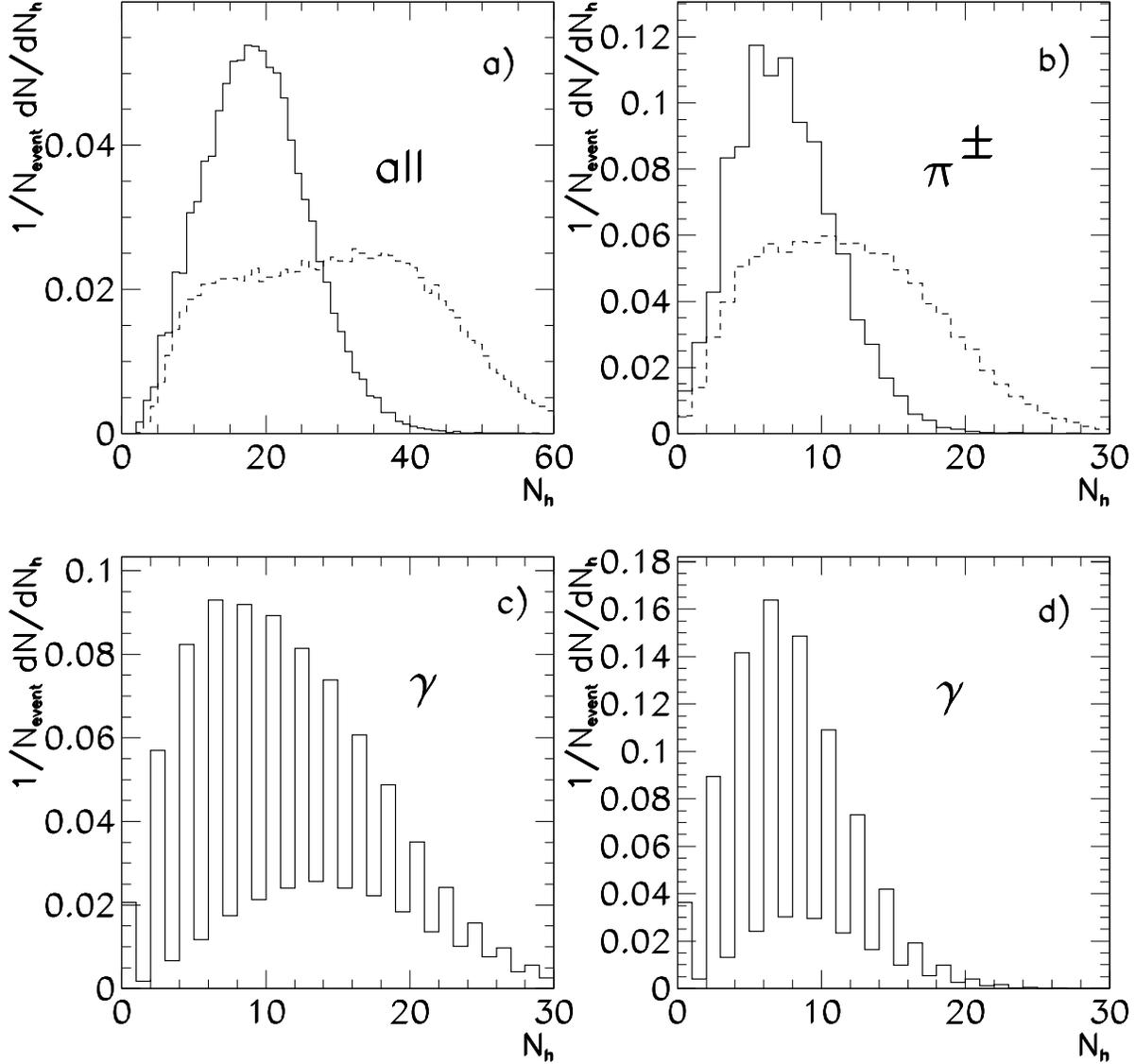,width=16.5cm,bbllx=20pt,
bblly=150pt,bburx=550pt,bbury=650pt,clip=}
\caption{\it Multiplicity distributions of (a) all stable particles, 
(b) charged pions from DIS on a $\pi^+$ (full curves) and DIS on a proton
(dashed curves). In (c) $\gamma$'s from DIS on the proton and 
(d) $\gamma$'s from DIS on $\pi^{+}$.}
\end{figure}

The flux factor given by Eq.~(\ref{splitpiN}) with a cut-off parameter
of the vertex form factor extracted
from the high-energy neutron production data \cite{HSS94} predicts
that the pion carries, on average, a fraction 0.3 of the proton beam
momentum \cite{HSS93}. This implies that as a first approximation
the pion-induced DIS processes can be viewed as an electron
scattering on the pion with effective energy
$E_{eff} \approx 0.3 \cdot E_{beam}$.
Because of the smaller energy of the pion one could expect smaller
multiplicity of electron-pion DIS events in
comparison to those for the electron-proton DIS. In Fig.~7 we compare
the model predictions (without experimental acceptance effects) of the
multiplicity spectra for DIS on the proton with those on $\pi^{+}$. 

The multiplicities in DIS events on the pion is
noticeably smaller than in DIS events on the proton;
the average multiplicity is about 20 and 30, respectively.
The dominant contribution to the multiplicity spectra comes from charged pions 
(11.6 on the proton vs. 7.5 on the
$\pi^{+}$) and $\gamma$'s (14.1 on the proton vs. 8.0 on $\pi^{+}$).
The even-odd fluctuations of the multiplicity spectra of photons
is not statistical, but caused mainly by the decay 
$\pi^{0} \rightarrow \gamma \gamma$.
Thus, as expected the multiplicity of standard DIS events is typically 
larger than in pion-induced DIS events. However, due to the large fluctuations 
in multiplicity and the overlap between the distributions for the two cases, 
as well as the distortions that limited experimental acceptance will create, 
it is not clear whether this difference can be used as a 
discriminator. This needs further considerations. 

\section{Conclusions}

The concept of a pion cloud in the nucleon was recently found to
be very useful \cite{HSS94,DYMCM} in understanding the Gottfried sum
rule violation observed by the New Muon Collaboration \cite{A91} and
the Drell-Yan asymmetry measured recently in the NA51 Drell-Yan
experiment at CERN \cite{B94}.
In the present paper we have investigated several quantities in
order to find useful observables which would help to verify this
concept using deep inelastic electron-proton scattering at HERA.
We have therefore analyzed the structure of deep inelastic
events induced by the pion-exchange mechanism. In particular,
we have studied distributions of final nucleons 
as well as rapidity and multiplicity spectra.

Most of the event characteristics do not provide a direct possibility
to distinguish the events from DIS on a pion from 
the ordinary events with DIS on a proton. 
A clear difference is, however, found in the energy
spectrum of outgoing neutrons.
We find that the pion cloud model predicts an energy distribution of
neutrons which substantially differs from the standard
hadronization models. While the pion-exchange mechanism leads to
an energy spectrum which peaks at an energy of about
$0.7 E_{beam}$, i.e. at about 500 GeV, the spectrum of neutrons
produced in the standard hadronization process following
DIS on the proton decreases monotonically with increasing
neutron energy. 
This should facilitate to discriminate between the two processes, 
in particular since they have cross sections of similar magnitude in this  
energy region. Therefore, the experiments with forward neutron 
calorimeters should shed new light on the nucleon structure in terms
of a pion content.  

We have shown that the pion cloud
induced mechanism practically does not contribute to the large rapidity
gap events observed recently by the ZEUS and H1 collaborations
\cite{lrg_ZEUS,lrg_H1}
and cannot be a severly competing mechanism for the pomeron exchange.
The multiplicity of the pion cloud induced events is about 60-70\%
of that for standard hadronization on the proton, but given the large 
fluctuations it is not clear to what extent this difference can be 
exploited. 

Our results on the pion exchange mechanism are more general than the 
detailed formulation of the pion cloud model. Since essentially the 
same pion flux factor is obtained in Regge phenomenology, our results
may also be taken as a representation of Regge-based expectations. 

In this study we have omitted experimental effects due to
finite appertures, clustering effects in the main detector,
finite energy thresholds, detector efficiencies, etc., which may distort
the observed spectra. Many of them are quite important in order to
understand and interpret the observed spectra and we plan a 
future study \cite{PS96} of such effects.

%---------------------------------------------------------------------

\end{document}